# Scattering of electrons by ionized impurities in semiconductors: quantum-mechanical approach to third body exclusion

## Dmitry Pozdnyakov

*Faculty of Radiophysics and Computer Technologies, Belarusian State University,*
*Nezavisimosty av.4, Minsk 220030, Belarus*
E-mail: pozdnyakov@tut.by

**Abstract:** As applied to the numerical simulation of electron transport and scattering processes in semiconductors an efficient model describing the scattering of electrons by the ionized impurities is proposed. On the example of GaAs at 77 and 300 K and Si at 300 K the dependencies of low-field electron mobility on the donor impurity concentration in the semiconductors are calculated in the framework of proposed model as well as in the framework of such most frequently used applied models as the Conwel-Weisskopf model and the Brooks-Herring one. After comparing the calculation results with the well-known experimental data it has been ascertained that the best agreement between the theory and experiment is achieved with application of the proposed model.

**Keywords:** ionized impurity; scattering; electron mobility; Monte Carlo simulation

# 1 Introduction

A lot of papers are devoted to the investigation of electron scattering by impurity ions in semiconductors due to particular importance of the issue. The studies of Conwel and Weisskopf [1], Brooks [2] and Herring (his work is unpublished), as well as Ridley [3] have already become the classical research works. The study of Chattopadhyay and Queisser [4], for example, can be considered as the review article on the issue. The process of electron scattering by impurity centers in semiconductors has been considered in detail in Ridley's monograph [5].

However it is necessary to note that in spite of the great attention to the issue from both the theoretical and practical points of view it is still far from its final solution. This is indicated by the results of comparison of experimental data with theoretical calculations obtained by a number of researchers (see, for example, [6–10]). Therewith, it should be added that the correct description of the electron scattering by ionized impurities has a significant applied importance equally with a fundamental one due to the intensive development of numerical simulation methods of charge carrier transport and scattering phenomena in semiconductors and semiconductor devices including the Monte Carlo method [11–14]. In this connection the purpose of the present work is the development of an effective applied model appropriately describing the scattering of electrons by ionized impurity atoms in semiconductors.

# 2 Theory

## 2.1 Basic calculations

Let us consider rather a general case of elastic scattering of free electrons by several different kinds of impurity atoms "$j$" all of which are the single-charged ions with the relative charge $Z_j = \pm 1$ located in the semiconductor crystal lattice sites randomly and independently in such a way that a distance between any two nearest charged impurities is greater than a distance between two nearest semiconductor atoms. For example, if the semiconductor is silicon that is doped with different donors and acceptors which are P, As, B and Ga then $j = 1$ for $P^+$, $j = 2$ for $As^+$, $j = 3$ for $B^-$ and $j = 4$ for $Ga^-$, respectively. It is obvious that both free electrons and free holes are scattered by the aggregate of $P^+$, $As^+$, $B^-$ and $Ga^-$ impurities.

As is known the relaxation processes in electron gas are mainly determined by the binary collisions of electrons with various scatterers [5]. Therefore while considering the scattering of charge carriers by the donor or acceptor ionized impurities it is necessary to calculate the scattering rate just in the binary collision approximation. As a result, taking into account the long-range character of the Coulomb potential, the mentioned necessity reduces to the problem of third body exclusion [5, 10]. The rigorous solution of such a problem in the framework of quantum-mechanical interpretation of the elastic binary collisions is the direct calculation of electron wave scattering amplitude by some charged impurity centre at a distance equal to a half of the distance between the centre and the nearest neighbour ion with a further averaging of the calculated amplitude over all the impurity centers. But from a practical point of view the rigorous solution of the problem is only methodologically interesting because of its obvious extremal computational complexity. Such a computational complexity nullifies the possibility of application of the calculated amplitude in further calculations of various parameters characterizing the electrical properties of semiconductors and in the simulation of charge carrier transport and scattering processes in semiconductors and semiconductor devices [6–8, 11–15]. Therefore the indicated problem is always solved in the framework of various approximations. For example, the simplest way of its solution has been proposed by Conwel and Weisskopf [1]. Its essence is to limit the impact parameter or, what is the same, to limit the minimal angle of scattering. A more rigorous probabilistic approach to the approximate solution of the third body exclusion problem has been proposed by Ridley [3]. To describe the collision of two charged particles Ridley used the classical mechanics formalism. His main obtained result is the following [3, 5]: the desired value of differential scattering cross-section in the classical approximation of binary collisions equals the product of Ridley's weighting factor and the value of differential cross-section of multiparticle scattering. There are no limitations for calculation of the initial value of differential cross-section of multiparticle scattering. It can be calculated in the framework of either classical or quantum-mechanical description of the multiparticle scattering. Further, similarly to Ridley, for an approximate solution of the third body exclusion problem let us apply the statistical (probabilistic) approach. But let us consider the scattering process from the viewpoint of quantum-mechanical description of binary collisions instead of the classical description.

In the case of scattering of electrons by charged centers the differential scattering cross-section per unit solid angle can be rigorously calculated by means of formula [16, 17]

$$\frac{dS_\Sigma}{d\Omega} = \sigma_\Sigma = |A_\Sigma|^2 = \left(\frac{m_e}{2\pi\hbar^2}\right)^2 \left|\langle\varphi'|V|\Psi^+\rangle\right|^2, \quad (1)$$

where $S_\Sigma$ is the total scattering cross-section, $A_\Sigma$ is the scattering amplitude, $m_e$ is the electron mass, $\hbar$ is the Planck constant, $\varphi = \varphi(\mathbf{r})$ is the wave function describing the plane wave, $\Psi^+ = \Psi^+(\mathbf{r})$ is the part of wave function $\Psi = \Psi^+ + \Psi^-$ being the general solution of the Schrödinger equation with the potential energy operator $V = V(\mathbf{r})$ describing the interaction of an electron with the aggregate of charged centers [16, 17]. $\Psi^+$ and $\Psi^-$ characterize respectively the wave outgoing from the scattering centers and the wave incoming to them. In eq. (1) " ′ " denotes the electron final state with the wave vector $\mathbf{k}'$. At that the initial electron state is characterized by the wave vector $\mathbf{k}$ ($|\mathbf{k}| = |\mathbf{k}'| = k$, $\mathbf{q} = \mathbf{k}' - \mathbf{k}$, $|\mathbf{q}| = q = 2k\sin(\theta/2)$, $\theta$ is the polar angle of scattering [5, 11–17]).

To pass from description of electron scattering in the free space to description of its scattering in the condensed matter (in the semiconductor in our case) with consideration of nonparabolicity (quasi-relativistic) and anisotropy effects [11–14] it is necessary to replace properly in eq. (1) the electron mass $m_e$ by the electron effective masses $m_c$ and $m_d$ taking into account an increase of them in $(1+2\eta E)$ times (see Herring and Vogt transformation in [11, 12] and section 8.2 "Relativistic Generalization" in [17]). Here $m_c$ and $m_d$ are the conductivity and density-of-states electron effective masses respectively ($3m_c^{-1} = m_l^{-1} + 2m_t^{-1}$ and $m_d^3 = m_l m_t^2$ [11–13]), $E$ is the electron kinetic energy, $\eta$ is the non-parabolicity parameter. After appropriate replacements we have the equality

$$\frac{dS_\Sigma}{d\Omega} = \sigma_\Sigma = \sqrt{\frac{m_c}{m_d}}|A_\Sigma|^2 = \sqrt{\frac{m_c}{m_d}}\left(\frac{m_d(1+2\eta E)}{2\pi\hbar^2}\right)^2 \left|\langle\varphi'|V|\Psi^+\rangle\right|^2. \quad (2)$$

In eq. (2) the operator $V$ is defined by the equality [15]

$$V(\mathbf{r}) = -\frac{e^2}{4\pi\varepsilon\varepsilon_0}\sum_{i,j}\frac{Z_j}{|\mathbf{r}-\mathbf{r}_{ij}|}, \quad (3)$$

where $e$ is the electron charge, $\varepsilon_0$ is the dielectric constant, $\varepsilon$ is the static relative dielectric constant of semiconductor, $\mathbf{r}_{ij}$ is the radius-vector determining the location of $i$-th charged impurity atom of $j$-th kind ($ij$-th charged impurity) in the space.

## 2.2 Separation of Coulomb potential

Now let us express $V(\mathbf{r})$ in the following form

$$V(\mathbf{r}) = \langle V(\mathbf{r}) \rangle + \Delta V(\mathbf{r}) = -\frac{e^2}{4\pi\varepsilon\varepsilon_0}\left(\sum_{i,j} Z_j \frac{1-\exp(-a^{-1}|\mathbf{r}-\mathbf{r}_{ij}|)}{|\mathbf{r}-\mathbf{r}_{ij}|} + \sum_{i,j} Z_j \frac{\exp(-a^{-1}|\mathbf{r}-\mathbf{r}_{ij}|)}{|\mathbf{r}-\mathbf{r}_{ij}|}\right), \quad (4)$$

where $a = (4\pi n_\Sigma/3)^{-1/3}$, $n_\Sigma = \sum_j n_j$, $n_j$ is the concentration of $j$-th kind ionized impurity atoms.

Such an expression of the potential energy $V$ allows it to be represented in the most separable way. Namely, while applying elementary numerical calculations, one can be convinced that $\langle V \rangle$ describes quite a smoothly varying macroscopic potential which almost coincides up to a constant factor with the electrostatic potential obtained by means of solution of the Poisson equation for the case of continuous distribution of the electrical charge in the space [12, 15]. And $\Delta V$ describes the microscopic deviation of $V$ from $\langle V \rangle$ due to a discrete distribution of the point-like ionized impurities in the space. Here it should be noted that the choice of exponent in eq. (4) is not occasional. The fact is that the only exponential function allows both $\langle V \rangle$ to be quite smooth and $\Delta V$ to be quite small in comparison with $\langle V \rangle$ at a distance $L/2$ ($L = n_\Sigma^{-1/3}$ is the average distance between the ionized impurities) from any charged impurity atom. In this case $\Delta V$ can be naturally considered as the perturbation potential to the potential $\langle V \rangle$, and we can calculate the scattering cross-section $\sigma$ by the method of successive approximations in framework of quantum perturbation theory. Moreover, exactly the exponential factor always appears in expressions for calculation of scattering cross-section or scattering rate in the case of consideration not only of the screening effects [2–5, 18] but also of other effects concerned with the electron localization due to the decoherence processes [18, 19] since, as is known, any incoherent scattering process can be considered as the process of quantum measurement [19].

Now if we suppose that $\Delta V(\mathbf{r}) \equiv 0$ then Eq. (2) obviously leads to the results almost coinciding with the result that can be obtained by substituting a wave function $\psi^+$ in eq. (2), where $\psi^+$ is the wave function being the solution of the Schrödinger equation [16, 17] with the potential energy operator determined by the Poisson equation [12, 15] with the continuous distribution of electric charge in the space. That is, if $\Delta V(\mathbf{r}) \equiv 0$ then the corresponding wave function $\Psi^+ \approx \psi^+$ characterizes the quantum average electron trajectories (the classical trajectories in the limit of very smoothly varying potential energy $\langle V \rangle$) in the macroscopic electric field created by the ionized impurities. In turn, if we assume that $\langle V(\mathbf{r}) \rangle \equiv 0$ then Eq. (2) will describe nothing else but the local two-particle interaction between an electron and the ionized impurity atom nearest to it [10] since the electrostatic potential of an ionized impurity atom is mainly localized in the region of volume $4\pi a^3/3$ around it. Thus, Eq. (2) determines nothing else than the differential scattering cross-section of electrons with quantum average trajectories of their motion in the macroscopic electric field and random deviations of their motion from the average trajectories because of scattering of charge carriers by the potential $\Delta V/e$.

During simulation of electron transport and scattering in semiconductors and semiconductor devices by quantum or classical Monte Carlo method the Wigner [20] or Boltzmann transport equation [11–15, 20], respectively, is always self-consistently solved along with the Poisson equation containing the continuous density of electric charge. Therefore to simulate properly the electron transport and scattering it is necessary to calculate the electron scattering cross-section (rate) just by the fluctuational potential $\Delta V/e$. As a result, in Eq. (2) the term containing the operator $\langle V \rangle$ we must omit and rewrite the equality in the form of the following expression

$$\sigma_{\text{total}} = a^2 \sqrt{\frac{m_c}{m_d}} \left(\frac{E_a^{\text{el}}}{E_a^{\text{qw}}}\right)^2 \left|\frac{1}{4\pi a^2} \sum_{i,j} Z_j \langle \varphi' | |\mathbf{r}-\mathbf{r}_{ij}|^{-1} \exp(-a^{-1}|\mathbf{r}-\mathbf{r}_{ij}|) | \psi^+ \rangle \right|^2, \quad (5)$$

where

$$E_a^{\text{qw}} = \frac{\hbar^2}{2m_d(1+2\eta E)a^2}, \quad (6)$$

$$E_a^{\text{el}} = \frac{e^2}{4\pi\varepsilon\varepsilon_0 a}. \quad (7)$$

## 2.3 The first Born approximation

Let us simplify Eq. (5) taking into account the fact that $\mathbf{r}_{ij}$ are the random independent vectors because the ionized impurities are located in the space randomly and independently. To do this, we shall apply the first Born approximation [15–17, 21]. In such a case one can obtain the following relations (see paragraph 5 "The scattering of charge carriers by impurity atoms" of section XIV and Appendix XIII "Averaging over the coordinates of impurity atoms" in [15])

$$\sigma_{\text{total}}^{\text{1stBA}}(q) = a^2 \sqrt{\frac{m_c}{m_d}} \left(\frac{E_a^{\text{el}}}{E_a^{\text{qw}}}\right)^2 \left|\frac{1}{4\pi a^2}\sum_{i,j} Z_j \left\langle \varphi' \left| |\mathbf{r}-\mathbf{r}_{ij}|^{-1} \exp(-a^{-1}|\mathbf{r}-\mathbf{r}_{ij}|) \right| \varphi \right\rangle \right|^2 = \sigma_{\text{incoh}}^{\text{1stBA}}(q) + \sigma_{\text{coh}}^{\text{1stBA}}(q), \quad (8)$$

$$\sigma_{\text{incoh}}^{\text{1stBA}}(q) = \sum_j \sigma_{\text{incoh},j}^{\text{1stBA}}(q) = \sum_j N_j a^2 \sqrt{\frac{m_c}{m_d}} \left(\frac{Z_j E_a^{\text{el}}}{E_a^{\text{qw}}}\right)^2 \left|\frac{1}{4\pi a^2} \left\langle \varphi' \left| r^{-1} \exp(-a^{-1} r) \right| \varphi \right\rangle\right|^2 =$$

$$N_\Sigma a^2 \sqrt{\frac{m_c}{m_d}} \left(\frac{E_a^{\text{el}}}{E_a^{\text{qw}}}\right)^2 \frac{1}{\left(1+q^2 a^2\right)^2}, \quad (9)$$

$$\sigma_{\text{coh}}^{\text{1stBA}}(q) = \begin{cases} 4\pi(N_\Sigma - 1)\sigma_{\text{incoh}}^{\text{1stBA}}(0), & q = 0, \\ 0, & q \neq 0, \end{cases} \quad (10)$$

where $N_\Sigma \to \infty$ is the total number of ionized impurities in the considered system [15]. It is seen from Eqs. (8) – (10) that the term describing the coherent scattering of an electron [4, 7, 8, 15, 21] by conjunction of all the ionized impurities vanishes for any value of $q$ except one $q = 0$ (the isotropic elastic scattering at $k = 0$, and the forward elastic scattering at $k \neq 0$). At that, as is well known, there is no need to take into account the electron scattering with such a dependence of $\sigma$ on $k$ and $\theta$ so that to describe correctly the electron transport phenomena [15]. Thus, we must take into account only the term describing the incoherent scattering of electrons by the ionized impurities. As a result the differential scattering cross-section of an electron by the ionized impurities per one impurity atom of $j$-th kind [15] can be calculated in the first Born approximation applying the formulae [16]

$$\sigma_j^{\text{1stBA}} = \frac{\sigma_{\text{incoh},j}^{\text{1stBA}}}{N_j} = \sqrt{\frac{m_c}{m_d}} \left|\frac{\lambda_{a,j}}{4\pi a} \left\langle \varphi' \left| r^{-1} \exp(-a^{-1} r) \right| \varphi \right\rangle\right|^2 = \sqrt{\frac{m_c}{m_d}} \left(\frac{\lambda_{a,j}}{q}\int_0^\infty \exp(-t)\sin(qat)\,dt\right)^2 =$$

$$a^2 \sqrt{\frac{m_c}{m_d}} \cdot \frac{\lambda_{a,j}^2}{\left(1+q^2 a^2\right)^2} = a^2 \sqrt{\frac{m_c}{m_d}} \cdot \frac{\lambda_{a,j}^2}{\left(1+4k^2 a^2 \sin^2(\theta/2)\right)^2}, \quad (11)$$

$$\lambda_{a,j} = Z_j E_a^{\text{el}} / E_a^{\text{qw}}. \quad (12)$$

The total scattering cross-section $S$ [5, 22] and the total momentum transfer cross-section $\bar{S}$ [23] are related to the differential scattering cross-section $\sigma$ by the well-known equalities

$$S(k) = 2\pi \int_0^\pi \sigma(k,\theta)\sin\theta\,d\theta = 4\pi \int_0^1 \sigma(k,t)\,dt, \quad (13)$$

$$\bar{S}(k) = 2\pi \int_0^\pi \sigma(k,\theta)(1-\cos\theta)\sin\theta\,d\theta = 8\pi \int_0^1 \sigma(k,t)t\,dt, \quad (14)$$

where $t = \sin^2(\theta/2)$.

It is known (see, for example, [16]) that the first Born approximation describes the single scattering event which is a jumplike quantum transition from one quantum state to another. In such a case the transition from Eq. (2) to Eq. (11), taking into account that $a \approx L/2$ (the strong localization of the electrostatic potential in the vicinity of the ionized impurity atom creating this potential), is nothing else but the third body exclusion principle which is formulated in the framework of quantum-mechanical description of the binary collisions in the first Born approximation.

Here a very important remark should be made. As a matter of fact the electric charge of mobile charge carriers (electrons and holes) must be also taken into account in Eq. (4) since it always enters the Poisson equation [12, 15]. Moreover, according to [5] the consistent consideration of electron–electron and electron–hole scattering implies to do it just in the binary collisions approximation. In particular, in case of solution of a more general scattering problem we will have to take into consideration not only the immobile electric charge of impurities but also the mobile charge carriers. But as the task like that is not a purpose of the study, we must take into account only the influence of the mobile charge carriers on the scattering process of a test electron [19] by the ionized impurities. It is easy to do it by generalizing Eq. (4) proceeding from the fact that in that case it is necessary to describe the scattering of the test electron by the aggregate of all the ionized impurities, holes and the rest electrons in the approximation of binary collisions. A quite obvious result following from such a generalization is the equality

$$a = \left(4\pi(n_\Sigma + n_e + n_h)/3\right)^{-1/3}, \quad (15)$$

which defines the value of $a$ finally. In Eq. (15) $n_e$ and $n_h$ are the electron and hole concentrations, respectively. Exactly this expression must be substituted in Eqs. (4) – (14).

## 2.4 Angular reduction method

All further calculations concerning a more accurate description of the ionized impurity scattering of electrons will be carried out only in the framework of the considered approximation for the binary collisions. At that, by analogy with [7], the method of reduction of one angular dependence of differential scattering cross-section to another dependence, which is ultimately less cumbersome and more convenient for all further calculations, will be applied. In [7], in particular, the roughest reduction was applied. Namely, it was the reduction of acute-angle dependence of differential cross-section of the Coulomb scattering to the dependence of differential cross-section of the isotropic scattering ($\sigma(k,\theta) \to \sigma_{\text{reduced}}(k) = \overline{S}(k)/4\pi$). So, a more accurate functional dependency $\sigma_j(k,\theta)$ will be searched in the following form [8]

$$\sigma_j(k,\theta) = \sigma_j^{\text{1stBA}}(k,\theta) + \sum_i \Delta\sigma_{ij}(k,\theta),$$

where $\Delta\sigma_{ij}$ is an additional differential scattering cross-section to the differential scattering cross-section in the first Born approximation which is caused by some "$i$"-th physical effect. At that according to the method of angular reduction the following approximation (see Eqs. (5.2) and (5.4) in [8])

$$\sigma_j(k,\theta) \approx G_j(k) \cdot \sigma_j^{\text{1stBA}}(k,\theta) \quad (16)$$

for the searched function will be applied. That is, the angular dependence of $\Delta\sigma_{ij}$ will be reduced to the angular dependence of $\sigma_j^{\text{1stBA}}$ (a far less rough angular reduction in comparison with the angular reduction used in [7]).

In accordance with the proposed in [7] method for determination of function $G_j$ it is necessary to solve the equation

$$\overline{S}_j(k) = \overline{S}_j^{\text{1stBA}}(k) + \sum_i \Delta\overline{S}_{ij}(k) = G_j(k) \cdot \overline{S}_j^{\text{1stBA}}(k) \quad (17)$$

relative to this function. It follows from Eq. (17) that

$$G_j(k) = \frac{\overline{S}_j(k)}{\overline{S}_j^{\text{1stBA}}(k)} = 1 + \sum_i \frac{\Delta\overline{S}_{ij}(k)}{\overline{S}_j^{\text{1stBA}}(k)} = 1 + \sum_i \alpha_{ij}(k). \quad (18)$$

In most cases, which are important from the practical point of view, the inequality $\alpha_{ij} \ll 1$ is true [23]. Therefore the approximate equality

$$G_j(k) = 1 + \sum_i \alpha_{ij}(k) \approx \prod_i \left(1 + \alpha_{ij}(k)\right) = \prod_i \frac{\overline{S}_j^{1stBA}(k) + \Delta \overline{S}_{ij}(k)}{\overline{S}_j^{1stBA}(k)} = \prod_i g_j^{(i)}(k) \quad (19)$$

can be applied instead of Eq. (18). It considerably simplifies all calculations and allows any "$i$"-th physical effect to be taken into account separately and independently from others. It is evident that the fewer the number of the effects and the closer the values of $\alpha_{ij}$ to zero, the better the approximation (19).

**2.5 The second Born approximation and its adjustment**
Let us now take into consideration the double scattering of an electron by an ionized impurity atom (the second Born approximation [16, 17]) in addition to the single scattering of the electron by the atom (the first Born approximation [16, 17]). According to [8, 17] the amplitude of electron scattering by a screened Coulomb potential (the Yukawa potential) in the second Born approximation is determined in our designations by the relations

$$A_j^{2ndBA} = A_j^{(1)} + A_j^{(2)} = A_j^{1stBA} + A_j^{(2)} = a\frac{\lambda_{a,j}}{1+q^2a^2} + a\frac{\lambda_{a,j}^2}{qa\tau}\left[\arctg\left(\frac{qa}{2\tau}\right) + \frac{i}{2}\ln\left(\frac{\tau+qka^2}{\tau-qka^2}\right)\right], \quad (20)$$

$$\tau = \sqrt{1+4k^2a^2+q^2k^2a^4} = \sqrt{1+4k^2a^2+4k^4a^4\sin^2(\theta/2)}. \quad (21)$$

As usually (see, for example, [21] and Paragraph 10.3.5 "An Example (the Yukawa Potential)" in [17]), let us omit the imaginary part of the scattering amplitude in Eq. (20). The reasons for this are as follows: (I) if $2ka \ll 1$ then $\operatorname{Im} A_j^{(2)} \ll \operatorname{Re} A_j^{(2)}$ and the imaginary term can be neglected; (II) if $2ka \geq \lambda_{a,j} \gg 1$ (the first Born approximation may be applied to the Yukawa potential [16]) then $\operatorname{Im} A_j^{(2)} \sim A_j^{(1)} \gg \operatorname{Re} A_j^{(2)} \Rightarrow A_j^{2ndBA} \not\to A_j^{1stBA}$, i.e. at $2ka \geq \lambda_{a,j} \gg 1$ the imaginary part of $A_j^{(2)}$ significantly degrades the convergence of the considered Born series and therefore it must be cast out. As a result, the differential scattering cross-section for the Yukawa potential in the second Born approximation can be calculated by the formula [17]

$$\sigma_j^{2ndBA}(q) = \sqrt{\frac{m_c}{m_d}}\operatorname{Re}^2\left(A_j^{2ndBA}(q)\right) = a^2\sqrt{\frac{m_c}{m_d}}\left(\frac{\lambda_{a,j}}{1+q^2a^2} + \frac{\lambda_{a,j}^2}{qa\tau}\arctg\left(\frac{qa}{2\tau}\right)\right)^2. \quad (22)$$

According to Eqs. (14), (18), (19), (21) and (22)

$$g_j^{2ndBA} = \frac{\overline{S}_j^{2ndBA}}{\overline{S}_j^{1stBA}} = \left(\int_0^1 \frac{t}{(1+4k^2a^2t)^2}dt\right)^{-1} \times$$

$$\int_0^1 \left(\frac{\sqrt{t}}{1+4k^2a^2t} + \frac{\lambda_{a,j}}{2}\cdot\frac{\arctg\left(ka(1+4k^2a^2+4k^4a^4t)^{-1/2}t^{1/2}\right)}{ka(1+4k^2a^2+4k^4a^4t)^{1/2}}\right)^2 dt. \quad (23)$$

Since there is an integral not represented by quadratures in the right part of Eq. (23), therefore similarly to [8] (see Eqs. (5.2), (5.4) and (5.5)) let us apply quite a simple approximation

$$g_j^{2ndBA} \approx \left\{1 + \frac{\lambda_{a,j}}{2}\left[1 + \frac{f_a}{4}\left(1 + \frac{1}{2+\ln(1+f_a/16)}\right)\right]\right\}^{-1\,2}, \quad (24)$$

which has been found by means of numerical methods. It approximates the right part of Eq. (23) with an exceptionally high degree of precision for any values of $ka$ and $\lambda_{a,j}$. Here the designation

$$f_a = 4k^2a^2 = 4\frac{m_c}{m_d}\cdot\frac{E(1+\eta E)}{E_a^{qw}(1+2\eta E)} \quad (25)$$

is introduced. It is very convenient for further calculations. The right part of Eq. (25) is a result of application of Eq. (6) and the dispersion relation for electrons in its usual form [11, 12, 14]

$$\frac{\hbar^2 k^2}{2m_c} = E(1+\eta E). \quad (26)$$

Let us adjust the function $g_j^{2ndBA}$ for the case when neither the first Born approximation nor the second one may not be applied for the Yukawa potential ($\lambda_{a,j} \gg 1 \gg 2ka$) [16]. To do it we must achieve the concordance between the obtained result, according to that (Eqs. (13), (14), (16), (19), (22) – (25)) $\bar{S}_j(0) = S_j(0) = 4\pi\sigma_j(0) = 4\pi m_c^{1/2} m_d^{-1/2} a^2 / c_j$, where $c_j = \left(\lambda_{a,j} + \lambda_{a,j}^2/2\right)^{-2}$ is a constant, and the result obtained with application of the rigorous phase-shift calculation of the scattering cross-section values for the Yukawa potential. The rigorous calculation gives $c_j \geq 1/4$ at $ka \to 0$ and $\lambda_{a,j} \to \infty$ [23, 24]. Also the obtained result must be in concordance with another result: $c_j \leq 4$ at $ka \to 0$ and $\lambda_{a,j} \to \infty$. It follows from a rather general semiclassical approach of Poklonski et al. (see [10] and references therein) in the framework of which the binary collisions between the particles are considered from the viewpoint of finite time of interaction of an electron with the Coulomb potential of each ionized impurity atom. Taking into consideration all of this along with the explicit form of Eq. (24), let us choose the adjusted function $\tilde{g}_j^{2ndBA}$ in the following form (see Eq. (13) in [24])

$$\tilde{g}_j^{2ndBA} = \frac{1}{1 - \xi(k)\lambda_{a,j} + c_j \xi^2(k)\lambda_{a,j}^2}. \quad (27)$$

Here $\xi(k)$ is such a function that relations $\xi(0) = 1$ and $\xi(k \to \infty) \to 0$ are true [24]. Let us determine the unknowns basing on the requirement of minimum discrepancy between $\tilde{g}_j^{2ndBA}$ and $g_j^{2ndBA}$ for all the values of $ka$ at such small values of $\lambda_{a,j}$ that the first Born approximation for the Yukawa potential may be applied ($\lambda_{a,j} \ll 1$) [16]. It is obvious that the minimum is reached when the relations

$$\left.\frac{\partial \tilde{g}_j^{2ndBA}(k)}{\partial \lambda_{a,j}}\right|_{\lambda_{a,j}=0} = \left.\frac{\partial g_j^{2ndBA}(k)}{\partial \lambda_{a,j}}\right|_{\lambda_{a,j}=0} \Rightarrow \xi(k) = \left[1 + \frac{f_a}{4}\left(1 + \frac{1}{2 + \ln(1 + f_a/16)}\right)\right]^{-1}, \quad (28)$$

$$\left.\frac{\partial^2 \tilde{g}_j^{2ndBA}(k)}{\partial \lambda_{a,j}^2}\right|_{\lambda_{a,j}=0} = \left.\frac{\partial^2 g_j^{2ndBA}(k)}{\partial \lambda_{a,j}^2}\right|_{\lambda_{a,j}=0} \Rightarrow 2(1-c_j)\xi^2(k) = \frac{1}{2}\left[1 + \frac{f_a}{4}\left(1 + \frac{1}{2 + \ln(1 + f_a/16)}\right)\right]^{-2} \quad (29)$$

are true. It immediately follows from them that $c_j = 3/4$ ($1/4 < 3/4 < 4$).

Taking into account Eqs. (11), (16) – (19), (22) – (29), it is easy to obtain the formulae for calculation of the differential scattering cross-section characterizing the binary collisions of electrons with the ionized impurities in the adjusted second Born approximation. With regard to further calculations it is convenient to present them in the form of

$$\sigma_j^{aa} = \sqrt{\frac{m_c}{m_d}}\left|\left\langle \varphi'\left|\frac{\Xi_{a,j}}{4\pi a}\cdot\frac{\exp(-a^{-1}r)}{r}\right|\varphi\right\rangle\right|^2 = q^{-2}\sqrt{\frac{m_c}{m_d}}\left(\Xi_{a,j}\int_0^\infty \exp(-t)\sin(qat)\,dt\right)^2 =$$

$$a^2\sqrt{\frac{m_c}{m_d}}\cdot\frac{\Xi_{a,j}^2}{\left(1+q^2a^2\right)^2} = a^2\sqrt{\frac{m_c}{m_d}}\cdot\frac{\Xi_{a,j}^2}{\left(1+f_a\sin^2(\theta/2)\right)^2}, \quad (30)$$

$$\Xi_{a,j} = \left\{\lambda_{a,j}^{-2} - \lambda_{a,j}^{-1}\left[1 + \frac{f_a}{4}\left(1 + \frac{1}{2 + \ln(1 + f_a/16)}\right)\right]^{-1} + \frac{3}{4}\left[1 + \frac{f_a}{4}\left(1 + \frac{1}{2 + \ln(1 + f_a/16)}\right)\right]^{-2}\right\}^{-1/2}. \quad (31)$$

## 2.6 Central-cell potential

Let us now consider one of the components of the central-cell potential [4–6, 8, 25–28]. But, in contrast to [6, 8], we will neglect the atomic core potential of the ionized impurity atom or, in other words, we will apply the isocoric approximation for the atomic core potential [5, 25–28] and take into account a more important potential of another nature. Its presence relates to the fact that the value of static relative dielectric constant of semiconductor $\varepsilon$, which is determined by the polarization of semiconductor atoms surrounding the impurity atom, is actually not a constant value but the value depending on the distance from the impurity atom (see [25–28] and Paragraph 4.4 "Central-cell contribution to charged impurity scattering" in [5]).

First, we will be considering an idealized case. Let the ionized impurity atom of $j$-th kind be surrounded by spherical wall with the radius $R_j$ coinciding with the effective atomic radius of the impurity. And let the wall be the absolutely impenetrable potential barrier for electrons which is an extremely thin and infinitely high. Also let the perfect polarizable continuous matter characterized by the value of static relative dielectric constant $\varepsilon$ be behind the wall. Then we have the obvious relations

$$F_j = \begin{cases} -\dfrac{Z_j e}{4\pi\varepsilon_0 r^2}, & r < R_j, \\ -\dfrac{Z_j e}{4\pi\varepsilon\varepsilon_0 r^2}, & r \geq R_j, \end{cases} = \begin{cases} -\dfrac{Z_j e}{4\pi\varepsilon\varepsilon_0 r^2} - \dfrac{(\varepsilon-1)Z_j e}{4\pi\varepsilon\varepsilon_0 r^2}, & r < R_j, \\ -\dfrac{Z_j e}{4\pi\varepsilon\varepsilon_0 r^2}, & r \geq R_j, \end{cases} =$$

$$-\dfrac{Z_j e}{4\pi\varepsilon\varepsilon_0 r^2} + \begin{cases} -\dfrac{(\varepsilon-1)Z_j e}{4\pi\varepsilon\varepsilon_0 r^2}, & r < R_j, \\ 0, & r \geq R_j, \end{cases} = F_j^0 + \Delta F_j$$

for the strength of electric field $F_j$ created by the ionized impurity atom with a charge $-Z_j e$. The electric field strength $F_j^0$ characterizes the potential energy of interaction of electrons with the ionized impurities in the matter with the static relative dielectric constant $\varepsilon$. The corresponding potential $\Delta V_j/e$ was already considered above. Therefore $\Delta F_j$ is nothing else but an additional strength of electric field determining the potential of central cell. After a simple integration of $\Delta F_j$ one can obtain the formula for calculation of the potential energy of interaction of an electron with the central cell:

$$\delta V_j = -\dfrac{(\varepsilon-1)Z_j e^2}{4\pi\varepsilon\varepsilon_0 r} \times \begin{cases} (1-rR_j^{-1}), & r < R_j, \\ 0, & r \geq R_j, \end{cases} \approx -\dfrac{Z_j e^2}{4\pi\varepsilon_0 r} \times \begin{cases} (1-rR_j^{-1}), & r < R_j, \\ 0, & r \geq R_j. \end{cases}$$

It is evident that for the region in the vicinity of the impurity atom ($r \ll R_j$) the inequality $\delta V_j/\Delta V_j > 1$ is true as always $\varepsilon > 1$. This means that that the central-cell potential is not weak, just the opposite, it is very strong but, at the same time, strongly localized. Taking into account the fact that $R_j < 0.2$ nm for any semiconductor, the inequality $2kR_j \ll 1$ is always true for every value of electron wave vector $k$ which is interesting from the viewpoint of simulation of electron transport and scattering. In this case mostly s-waves take part in the scattering processes and the first Born approximation may be applied [16, 24]. From the mathematical point of view all this is equivalent to the following chain of equalities [23, 24]:

$$\bar{S}_{\delta V} = S_{\delta V} = 4\pi\sigma_{\delta V}(0) = 4\pi\sqrt{\dfrac{m_c}{m_d}}\left(\dfrac{2m_d(1+2\eta E)}{\hbar^2}\int_0^\infty \delta V(r) r^2 dr\right)^2. \quad (32)$$

At that, because of peculiarities of scattering of electron s-waves (see [16]), the explicit form of the dependency of potential energy $\delta V$ on $r$ is not important since the electrons are scattered in the same manner for any other dependency of potential energy $\delta V^*$ on $r$ under the condition that [24]

$$\int_0^\infty \delta V(r) r^2 \mathrm{d}r = \int_0^\infty \delta V^*(r) r^2 \mathrm{d}r . \quad (33)$$

That is, the dependencies $\delta V(r)$ and $\delta V^*(r)$ are indistinguishable by the scattering of charge carriers if the single scattering of s-waves dominates.

Basing on the explicit form of the expressions determining $\Delta F_j$, it is easy to derive the formulae

$$\frac{\mathrm{d}Q_j}{\mathrm{d}r} = -\frac{(\varepsilon-1)Z_j e}{\varepsilon}\big(\delta_\mathrm{D}(r) - \delta_\mathrm{D}(r-R_j)\big),$$

$$\rho_j(r) = -\frac{(\varepsilon-1)Z_j e}{\varepsilon} \cdot \frac{\big(\delta_\mathrm{D}(r) - \delta_\mathrm{D}(r-R_j)\big)}{4\pi r^2}$$

describing the spatial distribution of corresponding effective electric charge $Q_j$ creating the central-cell potential. Here $\delta_\mathrm{D}$ is the Dirac delta function, $\rho_j$ is the effective charge density. It follows from the equalities that the absolutely impenetrable spherical wall is so charged that the total charge in the system is equal to zero. This screening surface charge on the wall is a result of polarization of the considered perfect continuous matter behind the wall.

Let us now consider the real case instead of the idealized one. In fact, on the one hand, the polarized matter is not continuous. It is sampled because of a discrete distribution of the atoms in the space. In addition, the real matter can not create a purely surface charge, the charge is always bulk. On the other hand, due to the chemical bounds between the atoms the potential barrier between any two nearest neighbors is low and easily penetrable. All of these must be taken into account. From the mathematical point of view it is elementary to do that. It is only necessary to replace the function $\delta_\mathrm{D}(r-R_j)/(4\pi r^2)$ by a smooth function of the screening charge density $\Omega_j(\mathbf{r})$ characterizing the spatial distribution of the electric charge in the vicinity of the ionized impurity atom. The rigorous calculation of the function $\Omega_j(\mathbf{r})$ implies much computational effort [25–28], besides it is not a purpose of this study. Therefore, as it is usually done (see, for example, [6, 8, 25, 27, 28]), we will apply a number of model simplifications and approximations. Let $\Omega_j(\mathbf{r}) = \Omega_j(r)$. Then we have the equality

$$\delta V_j(r) = -\frac{(\varepsilon-1)Z_j e^2}{4\pi\varepsilon\varepsilon_0 r}\left[1 - r\int_r^\infty\left(\int_0^y 4\pi t^2 \Omega_j(t)\mathrm{d}t\right)y^{-2}\mathrm{d}y\right] \quad (34)$$

for the central-cell potential. It has been obtained by integration of the Poisson equation in the spherical coordinates.

From the viewpoint of quantum-mechanical consideration of the screening effect a rather natural choice for $\Omega_j$ is a function describing the spatial distribution of the electric charge in the electron cloud of hydrogen-like atom in the ground quantum state. That is [6]:

$$\Omega_j(r) = \alpha_j^3 \exp(-\alpha_j r)/(8\pi), \quad (35)$$

where $\alpha_j \sim R_j^{-1}$ is a damping constant [25, 27, 28]. Then according to Eq. (34) we have formulae

$$\delta V_j(r) = -\frac{(\varepsilon-1)Z_j e^2}{4\pi\varepsilon\varepsilon_0 r}\big(1 + \alpha_j r/2\big)\exp(-\alpha_j r) \approx$$

$$-\frac{(\varepsilon-1)Z_j e^2}{4\pi\varepsilon\varepsilon_0 r}\exp(-\alpha_j^* r) = \delta V_j^*(r) , \quad (36)$$

$$\alpha_j^* = 2^{-1/2}\alpha_j . \quad (37)$$

It is easy to be convinced that Eq. (33) is true for $\delta V_j$ and $\delta V_j^*$ defined by Eqs. (36), (37). Here it should be noted that according to Eq. (34) the dependency $\delta V_j^*(r)$ is a result of integration of the dependency

$$\Omega_j^*(r) = (\alpha_j^*)^2 \exp(-\alpha_j^* r)/(4\pi r), \quad (38)$$

which was used in [8]. The spatial distribution of the screening charge described by Eq. (38) is the well-known result following from consideration of the screening effect in the framework of the semi-classical Thomas-Fermi model [8, 15, 16].

Basing on Eq. (36), the potential energy of interaction of the electron with the central cell can be represented as the following approximate equality

$$\delta V_j(r) \approx -\frac{(\varepsilon-1)Z_j e^2}{4\pi\varepsilon\varepsilon_0 r}\exp\left(-\frac{r}{c_j R_j}\right), \quad (39)$$

which is similar to Eq. (36) in [25]. Here $c_j = c(R_0, R_j) \sim 1$ is the coefficient depending on deformation of the semiconductor crystal lattice caused by difference between the effective atomic radius of the host material $R_0$ and the effective atomic radius of the dopant impurity $R_j$. It is evident that if the inequality $c_1 R_1 < c_2 R_2$ is true then the inverse inequality $\mu_1 > \mu_2$ is true for the corresponding values of electron mobility, what is confirmed by experimental data (see, for example, [8]). As rigorous calculations of the values of $c_j R_j$ are obviously time-consuming and, moreover, beyond the scope of the present study issue, let us apply the approximation of the isocoric impurity again. That is $c_j R_j \approx c_0 R_0 = \gamma^{-1} N_0^{-1/3}$, where $\gamma \sim 1$ is a constant, and $N_0$ is the intrinsic concentration of host material atoms. So, instead of the central-cell potential created by a specific kind of ionized impurity in a specific type of semiconductor we will consider the central-cell potential as if one of the semiconductor atoms is permanently charged with positive charge (the isocoric donor approximation) or negative charge (the isocoric acceptor approximation) [5, 25–28].

Thus, let us describe the potential energy of interaction of electron with the central cell by means of the following final expression

$$\delta V_j(r) = -\frac{(\varepsilon-1)Z_j e^2}{4\pi\varepsilon\varepsilon_0 r}\exp\left(-\gamma N_0^{1/3} r\right), \quad (40)$$

in which $\gamma$ is the fitting parameter. Proceeding from the fact that the main part of the screening charge must be on the atoms which are the nearest neighbors to the atom with the screened charge, the double inequality $1 < \gamma \leq 2^{1/2}$ can be easily obtained from Eqs. (35) – (40) (search of maximums of functions $4\pi r^2 \Omega_j^*(r)$ and $4\pi r^2 \Omega_j(r)$). The value refinement for $\gamma$ can be done in the framework of very simple geometrical considerations (see Appendix 4.10 "Average separation of impurities" in [5]). Basing on them, the estimation

$$\gamma = \gamma_0 = 2^{1/2}\Gamma(4/3) \approx 1.263 \quad (41)$$

can be obtained taking into account Eqs. (35) – (37), (40). Here $\Gamma$ is the Euler Gamma function.

To verify the calculations related to the central-cell potential let us now calculate the ionization energy of isocoric donor atom P in Si [5, 25–28] with consideration of the central-cell potential defined by Eq. (40). Proceeding from the equality, in the first approximation of the perturbation theory [16] for the hydrogen-like isocoric donor impurity [25–28] we have the formulae

$$E_I(\gamma) = E_I^0 + \Delta E_I^0(\gamma) = E_I^0 + 4\pi\int_0^\infty |\delta V_j(\gamma, r)|\phi^2(r)r^2 dr =$$

$$E_I^0 + 8|Z_j|(\varepsilon-1)E_I^0\int_0^\infty \exp\left(-\gamma a_B N_0^{1/3} t\right)\exp(-2t) t\, dt = \left[1 + \frac{8(\varepsilon-1)}{\left(2+\gamma a_B N_0^{1/3}\right)^2}\right]E_I^0, \quad (42)$$

$$E_I^0 = \frac{e^2}{8\pi\varepsilon\varepsilon_0 a_B}, \quad (43)$$

$$a_B = \frac{4\pi\varepsilon\varepsilon_0 \hbar^2}{m_c e^2} \quad (44)$$

for calculation of the ionization energy. Here $\phi$ is the radial wave function describing the electron in 1s state of the hydrogen-like atom. According to Eqs. (42) – (44) the ionization energy of P in Si is determined by the following values: $E_I(1) = 43.9$ meV, $E_I(\gamma_0) = 37.9$ meV, $E_I(2^{1/2}) = 35.7$ meV and $E_I(\infty) = 25.3$ meV (there is no central-cell potential). Although the calculation of $E_I$ by Eqs. (42) – (44) is rather a rough estimation, nevertheless the estimation results for $1 < \gamma \leq 2^{1/2}$ are in a good agreement with both the results of more rigorous calculations ($E_I = 42.4$–$44.8$ meV [26–28]) with application of the different model dependencies $\delta V(r)$ uniquely related to dependencies $\varepsilon^{-1}(r)$ (see [25, 27, 28]) and the results of experimental measurements ($E_I = 45.5$ meV [26–28]).

## 2.7 Central-cell scattering

If the only scattering potential was the central-cell potential of the ionized impurities located in the space randomly and independently then according to all discussed above we could immediately write the corresponding expressions for calculation of the differential scattering cross-section characterizing the binary collisions of electrons with the central cells of the ionized impurities in the adjusted second Born approximation:

$$\sigma_j^{bb} = \sqrt{\frac{m_c}{m_d}} \left| \left\langle \varphi' \left| \frac{\Xi_{b,j}}{4\pi b} \cdot \frac{\exp(-b^{-1}r)}{r} \right| \varphi \right\rangle \right|^2 = q^{-2} \sqrt{\frac{m_c}{m_d}} \left( \Xi_{b,j} \int_0^\infty \exp(-t)\sin(qbt)\,\mathrm{d}t \right)^2 =$$

$$b^2 \sqrt{\frac{m_c}{m_d}} \cdot \frac{\Xi_{b,j}^2}{\left(1+q^2b^2\right)^2} = b^2 \sqrt{\frac{m_c}{m_d}} \cdot \frac{\Xi_{b,j}^2}{\left(1+f_b \sin^2(\theta/2)\right)^2}, \quad (45)$$

$$\Xi_{b,j} = \left\{ \lambda_{b,j}^{-2} - \lambda_{b,j}^{-1} \left[ 1 + \frac{f_b}{4}\left(1 + \frac{1}{2 + \ln(1+f_b/16)}\right) \right]^{-1} + \frac{3}{4}\left[ 1 + \frac{f_b}{4}\left(1 + \frac{1}{2 + \ln(1+f_b/16)}\right) \right]^{-2} \right\}^{-1/2}, \quad (46)$$

$$f_b = 4k^2b^2 = 4\frac{m_c}{m_d} \cdot \frac{E(1+\eta E)}{E_b^{\mathrm{qw}}(1+2\eta E)}, \quad (47)$$

$$\lambda_{b,j} = Z_j E_b^{\mathrm{el}} / E_b^{\mathrm{qw}}, \quad (48)$$

$$E_b^{\mathrm{qw}} = \frac{\hbar^2}{2m_d(1+2\eta E)b^2}, \quad (49)$$

$$E_b^{\mathrm{el}} = \frac{(\varepsilon-1)e^2}{4\pi\varepsilon\varepsilon_0 b}, \quad (50)$$

$$b^{-1} = a^{-1} + \gamma_0 N_0^{1/3}. \quad (51)$$

## 2.8 Total scattering

As a matter of fact there is the Yukawa potential already discussed besides the screened central-cell potential. And the electron is scattered by this total potential. Therefore with the utmost rigor by analogy with Eq. (22) it should be written

$$\sigma_j^{\mathrm{sum}} = \sqrt{\frac{m_c}{m_d}} \operatorname{Re}^2\left( A_{j,aa}^{(1)} + A_{j,aa}^{(2)} + 2A_{j,ab}^{(2)} + A_{j,bb}^{(2)} + A_{j,bb}^{(1)} \right).$$

However in our case ($b \ll a \Rightarrow \operatorname{Re} A_{j,aa}^{(2)} \gg \operatorname{Re} A_{j,ab}^{(2)} \gg \operatorname{Re} A_{j,bb}^{(2)}$) it is quite sufficient to apply the following approximation (the part of the scattering amplitude describing the double scattering of electron initially by the Yukawa potential and then by the screened central-cell potential or initially by the screened central-cell potential and then by the Yukawa potential is omitted)

$$\sigma_j^{\mathrm{sum}} \approx \sqrt{\frac{m_c}{m_d}} \operatorname{Re}^2\left( A_{j,aa}^{(1)} + A_{j,aa}^{(2)} + A_{j,bb}^{(2)} + A_{j,bb}^{(1)} \right)$$

for the differential cross-section of electron scattering by the total potential. Such an approximation is very convenient for further calculations from the practical point of view. In particular, in contrast to the previous equality the application of the considered methods of angular reduction and adjustment of the second Born approximation to this approximate equality allows the result in the expected form of

$$\tilde{\sigma}_j^{sum} \approx \sqrt{\frac{m_c}{m_d}} \left| \left\langle \varphi' \left| \frac{\Xi_{a,j}}{4\pi a} \cdot \frac{\exp(-a^{-1}r)}{r} + \frac{\Xi_{b,j}}{4\pi b} \cdot \frac{\exp(-b^{-1}r)}{r} \right| \varphi \right\rangle \right|^2 =$$

$$q^{-2} \sqrt{\frac{m_c}{m_d}} \left( \int_0^\infty \left( \Xi_{a,j} \sin(qat) + \Xi_{b,j} \sin(qbt) \right) \exp(-t) \, dt \right)^2 =$$

$$\sqrt{\frac{m_c}{m_d}} \left( \frac{a\Xi_{a,j}}{1+q^2a^2} + \frac{b\Xi_{b,j}}{1+q^2b^2} \right)^2 = \sqrt{\frac{m_c}{m_d}} \left( \frac{a\Xi_{a,j}}{1+f_a \sin^2(\theta/2)} + \frac{b\Xi_{b,j}}{1+f_b \sin^2(\theta/2)} \right)^2 = \sigma_j^{aa} + \sigma_j^{ab} + \sigma_j^{bb}, \quad (52)$$

$$\sigma_j^{ab} = 2ab \sqrt{\frac{m_c}{m_d}} \cdot \frac{\Xi_{a,j}\Xi_{b,j}}{(1+q^2a^2)(1+q^2b^2)} = 2ab \sqrt{\frac{m_c}{m_d}} \cdot \frac{\Xi_{a,j}\Xi_{b,j}}{(1+f_a \sin^2(\theta/2))(1+f_b \sin^2(\theta/2))} \quad (53)$$

to be easily obtained. As would be expected, the differential cross-section of electron scattering by the total potential of the ionized impurities is the sum of three partial components. Namely, $\sigma_j^{aa}$ describes the incoherent scattering of electron by the Yukawa potential (as if there is no screened central-cell potential); $\sigma_j^{bb}$ describes the incoherent scattering of electron by the screened central-cell potential (as if there is no Yukawa potential); and lastly $\sigma_j^{ab}$ describes the coherent scattering of electron by the Yukawa potential and the screened central-cell potential.

**2.9 Overlap integral**
Let us finally take into account the fact that the overlap integral of the Bloch functions characterizing the initial and final states of electron is not a unit [11–14]. We will apply the standard expression for the square magnitude of the overlap integral from [14]. In our designations it can be represented in the form of

$$I_0(\theta) = 1 - \frac{4\eta E(1+\eta E)}{(1+2\eta E)^2} \sin^2(\theta/2). \quad (54)$$

Proceeding from Eqs. (52) – (54) we can immediately write the formula

$$\sigma_j^{int}(\theta) = I_0(\theta) \tilde{\sigma}_j^{sum}(\theta) = I_0(\theta) \left( \sigma_j^{aa}(\theta) + \sigma_j^{ab}(\theta) + \sigma_j^{bb}(\theta) \right). \quad (55)$$

The values of $I_0$ are very close to a unit for the values of $E$ which are interesting from the viewpoint of simulation of electron transport and scattering in semiconductors. Therefore let us reduce the angular dependence of $I_0$ on $\theta$. Applying Eqs. (16) – (19), (54) and (55), after simple calculations it is easy to obtain the final expression for calculation of the differential cross-section of electron scattering by the ionized impurities. That is

$$\sigma_j^{final}(\theta) = \tilde{\sigma}_j^{int}(\theta) = I_{aa} \sigma_j^{aa}(\theta) + I_{ab} \sigma_j^{ab}(\theta) + I_{bb} \sigma_j^{bb}(\theta), \quad (56)$$

where

$$I_{aa} = 1 - \frac{4\eta E(1+\eta E)}{(1+2\eta E)^2} \cdot \frac{f_a^2 + 2f_a - 2(1+f_a)\ln(1+f_a)}{f_a(1+f_a)\ln(1+f_a) - f_a^2}, \quad (57)$$

$$I_{ab} = 1 - \frac{4\eta E(1+\eta E)}{(1+2\eta E)^2} \cdot \frac{f_a^2(f_b - \ln(1+f_b)) - f_b^2(f_a - \ln(1+f_a))}{f_a f_b (f_a \ln(1+f_b) - f_b \ln(1+f_a))}, \quad (58)$$

$$I_{bb} = 1 - \frac{4\eta E(1+\eta E)}{(1+2\eta E)^2} \cdot \frac{f_b^2 + 2f_b - 2(1+f_b)\ln(1+f_b)}{f_b(1+f_b)\ln(1+f_b) - f_b^2}. \quad (59)$$

Due to the application of approximation of the isocoric impurity the subscript $j$ eventually enters Eq. (56) explicitly or implicitly only through the multiplier $Z_j = \pm 1$. Therefore there is no need to distinguish neither different donors between themselves nor different acceptors between themselves. It is enough only to distinguish them all with the sign of their electric charge. In particular, $j =$ "D" ($Z_D = +1$) for any donor and $j =$ "A" ($Z_A = -1$) for any acceptor. Thus, we achieve the universal description of electron scattering by the ionized donor impurities (attractive potential) and/or the ionized acceptor impurities (repulsive potential) [23, 24].

**2.10 Scattering rate**
The total cross-section of electron scattering by the ionized impurities $S_j^{\text{final}}$ can be calculated by Eq. (13). The electron scattering rate $W_j$ is determined by the total scattering cross-section $S_j^{\text{final}}$. It is expressed by the formula [8, 11, 12, 14, 22]
$$W_j(E) = v_g(E) S_j^{\text{final}}(E) n_j, \quad (60)$$
where $v_g$ is the electron group velocity, namely $v_g(E_0) = \langle |\nabla_{\mathbf{p}} E| \rangle_{\mathbf{p}:E(\mathbf{p})=E_0}$ [11, 12, 14, 15].

Eqs. (13), (30), (31), (45), (46), (53), (56) – (60) allow all the necessary expressions for calculation of the scattering rate of electrons $W_j$ by the $j$-th kind ionized impurities to be obtained in the explicit form. Namely, they are
$$W_j = W_j^{aa} + W_j^{ab} + W_j^{bb}, \quad (61)$$
$$W_j^{aa} = 4\pi a^2 n_j I_{aa} \left( \frac{2E(1+\eta E)}{m_d(1+2\eta E)^2} \right)^{1/2} \frac{\Xi_{a,j}^2}{1+f_a}, \quad (62)$$
$$W_j^{ab} = 8\pi ab n_j I_{ab} \left( \frac{2E(1+\eta E)}{m_d(1+2\eta E)^2} \right)^{1/2} \frac{\Xi_{a,j} \Xi_{b,j}}{f_a - f_b} \ln\left( \frac{1+f_a}{1+f_b} \right), \quad (63)$$
$$W_j^{bb} = 4\pi b^2 n_j I_{bb} \left( \frac{2E(1+\eta E)}{m_d(1+2\eta E)^2} \right)^{1/2} \frac{\Xi_{b,j}^2}{1+f_b}. \quad (64)$$

**2.11 Polar scattering angle**
To use the obtained results in the Monte Carlo simulation of electron transport and scattering in semiconductors it is also necessary to calculate the probability of electron scattering at a given angle $\theta_j$ [7, 11–14, 22]. But the necessity along with the explicit form of the obtained equalities reduces, after application of the direct technique [11, 12], to the need of solution of a transcendental equation every time after the electron–ion scattering. However, taking into account the statistical basis of the Monte Carlo method, the indicated problem can be avoided by the pure statistical way applying the rejection technique [11, 12]. In order to do this it is necessary to consider the incoherent scattering of electrons by the screened Coulomb potential, the incoherent scattering of electrons by the screened central cell potential and their coherent scattering by the screened Coulomb and screened central cell potentials as scattering mechanisms which are independent from each other like all other scattering mechanisms (phonons, plasmons, crystal lattice defects, etc.). In that case every such a considered scattering mechanism can be characterized by a scattering angle distribution function itself. Each corresponding partial scattering rate ($W_j^{aa}$, $W_j^{bb}$ and $W_j^{ab}$) along with the self-scattering one contribute to the total scattering rate $\Gamma$ [11–14]. As a result any such a scattering mechanism is selected using rejection technique or combined technique [11–14] in accordance with its statistical weight ($W_j^{aa}/\Gamma$, $W_j^{bb}/\Gamma$ and $W_j^{ab}/\Gamma$, respectively). In the framework of the proposed approach the expressions for selection of a random scattering angle $\theta_\rho$ can be easily obtained with application of the known mathematical methods [11–14], and represented as
$$\cos(\theta_\rho^{aa}) = 1 - 2\frac{1-\rho}{1+\rho f_a}, \quad (65)$$
$$\cos(\theta_\rho^{bb}) = 1 - 2\frac{1-\rho}{1+\rho f_b}, \quad (66)$$

$$\cos(\theta_\rho^{ab}) = 1 - 2\frac{(1+f_a)^\rho - (1+f_b)^\rho}{f_a(1+f_b)^\rho - f_b(1+f_a)^\rho}. \quad (67)$$

Here $\rho$ is the random number uniformly distributed between 0 and 1.

## 3 Results of calculations and discussion

As an example the results of calculation of the total electron scattering rate ($W = W_D + W_A$) according to Eqs. (61) – (64) for some different values of donor and acceptor concentrations $n_D$ and $n_A$, respectively, in Si are represented in Fig. 1. The figure reflects the well known result. Namely, electrons are always scattered by the attractive potential more intensively than by the repulsive one [4, 23, 24].

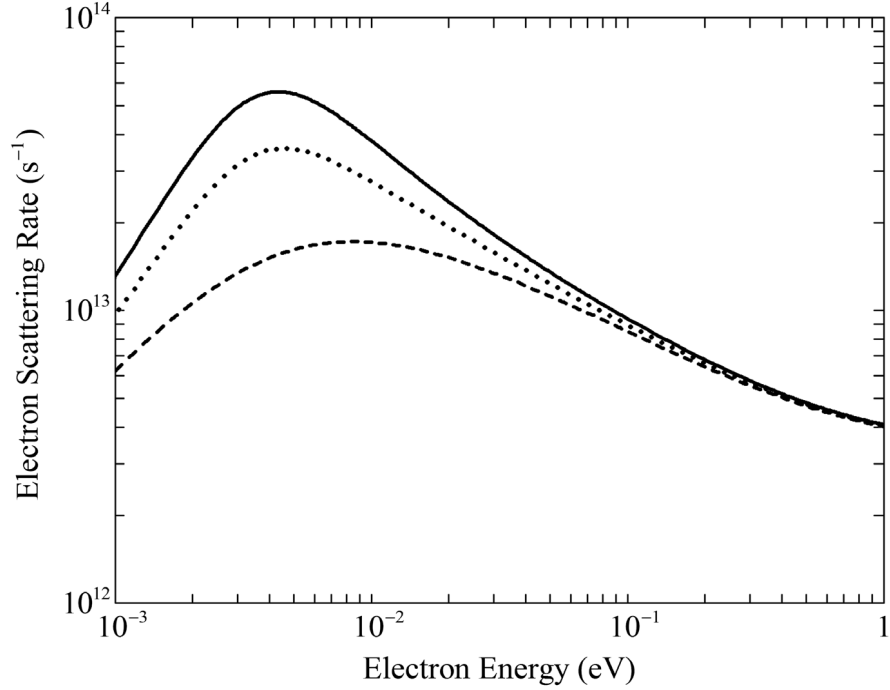

**Fig. 1** Electron scattering rates in Si calculated according to PM: $n_D = 10^{16}$ cm$^{-3}$ and $n_A = 0$ (n-type semiconductor, solid curve), $n_D = n_A = 5 \cdot 10^{15}$ cm$^{-3}$ (completely compensated semiconductor, dotted curve), $n_D = 0$ and $n_A = 10^{16}$ cm$^{-3}$ (p-type semiconductor, dashed curve).

To verify the proposed model (PM) describing the scattering of electrons by the ionized impurities the low-field mobility of electrons in GaAs and Si doped only with some donor impurities was calculated at different values of electric field strength $F_0$ by means of the Monte Carlo method with ensemble of $10^6$ particles. To do it the electron drift velocity was calculated during 120 ps through femtosecond interval after the electric field had been switched on. The time-averaged electron drift velocity was used in calculations of the electron low-field mobility. The time-averaging interval was the last 20 ps of simulation time. Then the results of calculation were compared with the well-known experimental data. While calculating the electron scattering rates, the Pauli exclusion principle (the electron gas degeneracy) was taken into consideration. In GaAs electron band structure Γ-valley in the spherical non-parabolic approximation for the isoenergetic surfaces and L-valleys in the ellipsoidal non-parabolic approximation for the isoenergetic surfaces, as well as X-valleys in the ellipsoidal non-parabolic approximation for the isoenergetic surfaces in Si electron band structure were taken into account [5, 11, 12, 14]. All the dominant electron scatterers in the semiconductors were considered along with the ionized impurity. In case of GaAs the electron scatterers of the following kind were taken into account: the polar optical phonons [12, 14], the acoustic phonons (inelastic scattering) [12, 14], the piezoelectric phonons (elastic and equipartition approximation) [5, 12, 14], the intervalley phonons [12, 13], and the plasmons [12]. The electron–plasmon coupling was considered in the framework of electron–electron model with application of the first order analytical approximation for the plasmon dispersion relation [29]. In case of Si the electron scatterers of the following kind were taken into account: the intravalley and intervalley acoustic (LA and TA) and optical (LO and TO) phonons in the framework of isotropic quadratic phonon dispersion model [30] as well as plasmons [12, 31] in the framework of electron–electron model with application of the first order analytical approximation for the plasmon dispersion relation [29]. The electron–electron scattering in the semiconductors was neglected since, as is well known [8, 11, 12, 32], the binary collisions of identical charge carriers almost do not influence their low-field mobility.

In Figs. 2–4 the experimental data ranges (EDRs) are shown along with the results of theoretical calculations.

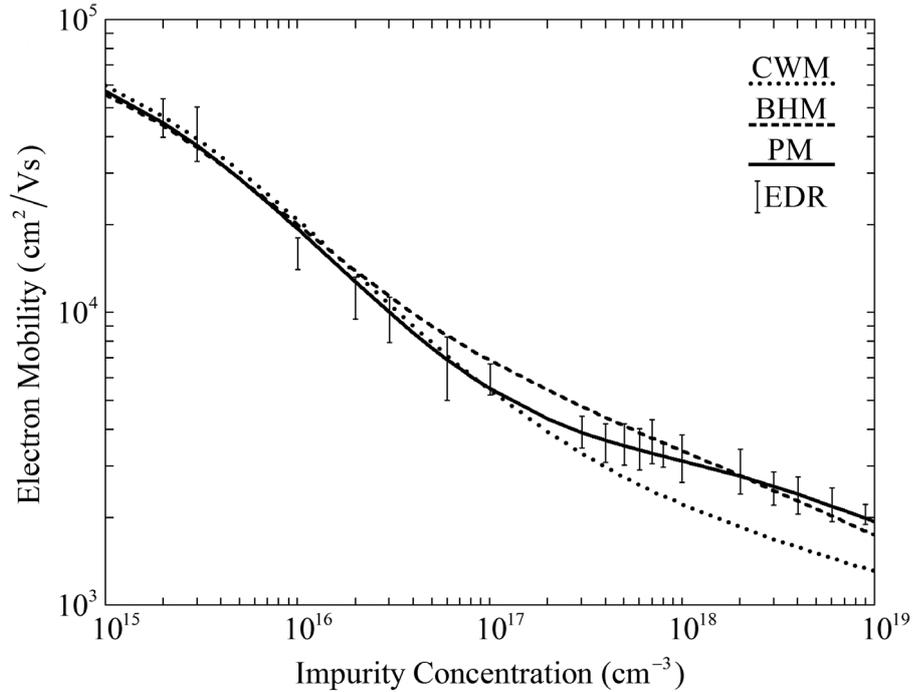

**Fig. 2** Low-field mobility of electrons in GaAs at $T = 77$ K: experimental data (EDR) and theoretical calculations with application of CWM, BHM and PM under the condition that $F_0 = 5$ V/cm.

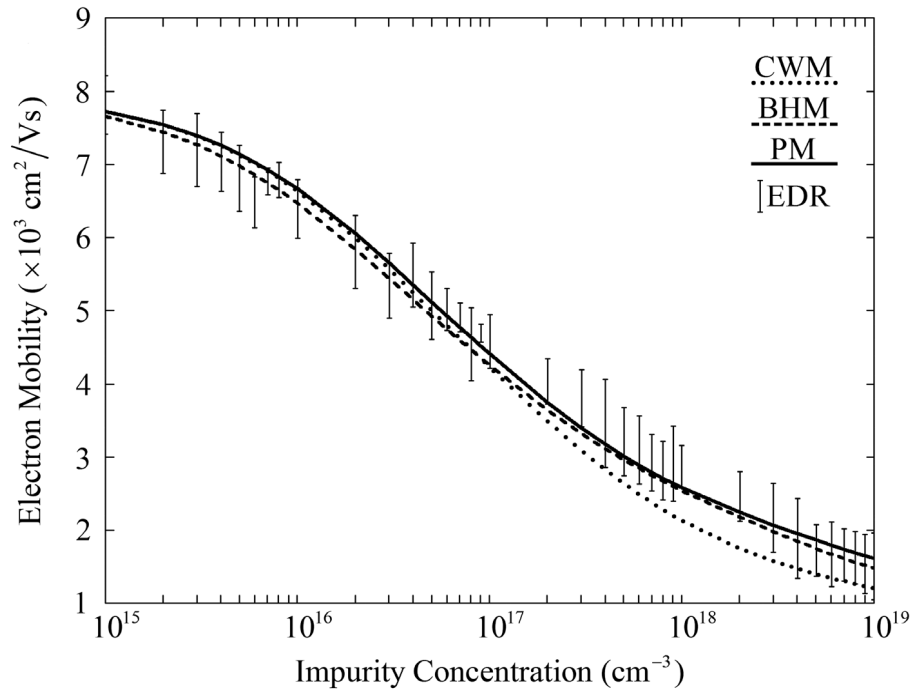

**Fig. 3** Low-field mobility of electrons in GaAs at $T = 300$ K: experimental data (EDR) and theoretical calculations with application of CWM, BHM and PM under the condition that $F_0 = 50$ V/cm.

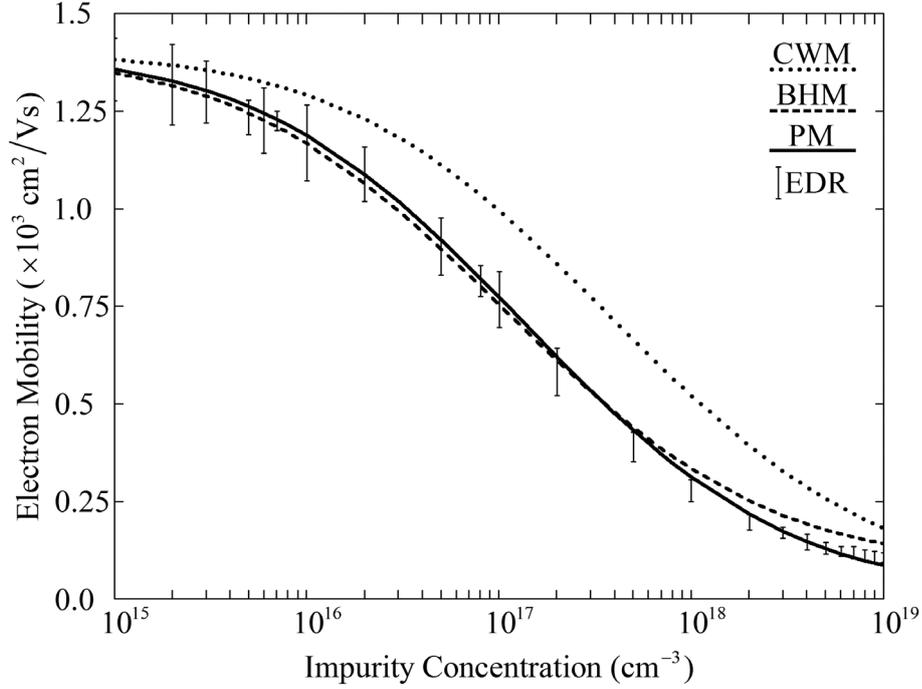

**Fig. 4** Low-field mobility of electrons in Si at $T = 300$ K: experimental data (EDR) and theoretical calculations with application of CWM, BHM and PM under the condition that $F_0 = 200$ V/cm.

In Figs. 2 and 3 the experimental data are averaged EDRs for the measured values of electron mobility from [6, 9, 21, 33–35], and in Fig. 4 the experimental data are averaged EDRs for the measured values of electron mobility from [33, 36–40]. The represented theoretical results include the calculations of low-field mobility of electrons in GaAs at temperatures $T = 77$ and 300 K as well as in Si at $T = 300$ K according to not only PM, but also two other models that are used in the Monte Carlo simulations in the most cases (see, for example, [7, 8, 11, 12, 14]). They are the Conwel–Weisskopf model (approach) CWM [1], according to which the scattering of an electron by the Coulomb potential of an ionized impurity is described in the framework of classical mechanics with elimination of the total scattering cross-section divergence by limiting the impact parameter value, and the Brooks–Herring model (approach) BHM [2], according to which the scattering of an electron by the screened Coulomb potential of an ionized impurity (it is believed that the Coulomb potential of ionized impurity is screened by the free charge carriers) is described in the framework of quantum mechanics. There are no results of calculation in the figures in accordance with yet another model. It is the (classical) Ridley model (approach) RM [3] that adjusts CWM with BHM. Sometimes RM is used in the Monte Carlo simulations but much less frequently than CWM or BHM [22]. Its application always leads to such values of electron low-field mobility $\mu_{RM}$ that almost coincide with the values of the mobility $\mu_{BHM}$ calculated according to BHM ($\mu_{RM} \gtrsim \mu_{BHM}$) [22]. RM is not so widespread as CWM or BHM because of a high computational complexity of the polar scattering angle $\theta$ selection (or impact parameter selection) [22]. For the same reason (a higher computational complexity) other models are not used in the Monte Carlo simulation of electron scattering by ionized impurities in semiconductors (see, for example, [5, 10] and references therein). Here some remarks should be added. In particular, the transition from PM to BHM is elementary from the mathematical point of view. Its essence is the replacement of the value of $a$ defined by Eq. (15) by the value of the Thomas–Fermi screening radius $r_{TF} = \beta_s^{-1}$ [4, 6–8, 31] in all the equalities above. At such a transition, in contrast to [6–8], there will be no additional terms in Eqs. (52), (55) and (56) characterizing two-ion coherent scattering of electrons. The fact is that, as it was shown earlier (see Eqs. (8) – (10)), when considering not only the two-ion coherent scattering but three-, four-, ... , N-ion coherent scattering in conjunction at $N \to \infty$ such a term vanishes for all the values of $q$ except one $q = 0$ in the case of a random and independent location of impurity atoms in the semiconductor crystal lattice [15]. And, as is well known, there is no need to take into account the scattering like that. The necessity of consideration of the coherent scattering along with the non-coherent one arises only if there is a partial ordering in a spatial location of the ionized impurity atoms [15, 35].

In Figs. 2–4 the calculated values of electron mobility are limited by the value of donor density $n_D = 10^{19}$ cm$^{-3}$. This is due to the fact that at higher dopant concentrations one can not hope to obtain the valid calculation results without consideration of a number of effects that are typical to the heavily doped semiconductors (impurity dressing effects [21]; tails of the density-of-states, hopping conduction, electron–hole droplets [15]; etc.). In Figs. 2 and 3 the theoretical curves were calculated taking into account the fact that only a part of the impurity atoms embeds into the GaAs crystal lattice sites in the donor state [41]. It was assumed that the scattering of electron by such a non-donor atom is the isotropic scattering of the charge carrier by the impenetrable sphere [42] with the effective total scattering cross-section value equal to $N_0^{-2/3}$.

In Fig. 5 the results of calculation of electron low-field mobility according to PM for different values of $\gamma$ are represented. It follows from Figs. 2–5 that the value $\gamma = \gamma_0$ ensures the best agreement between the theoretical calculations and experimental measurements for the electron mobility in GaAs and Si.

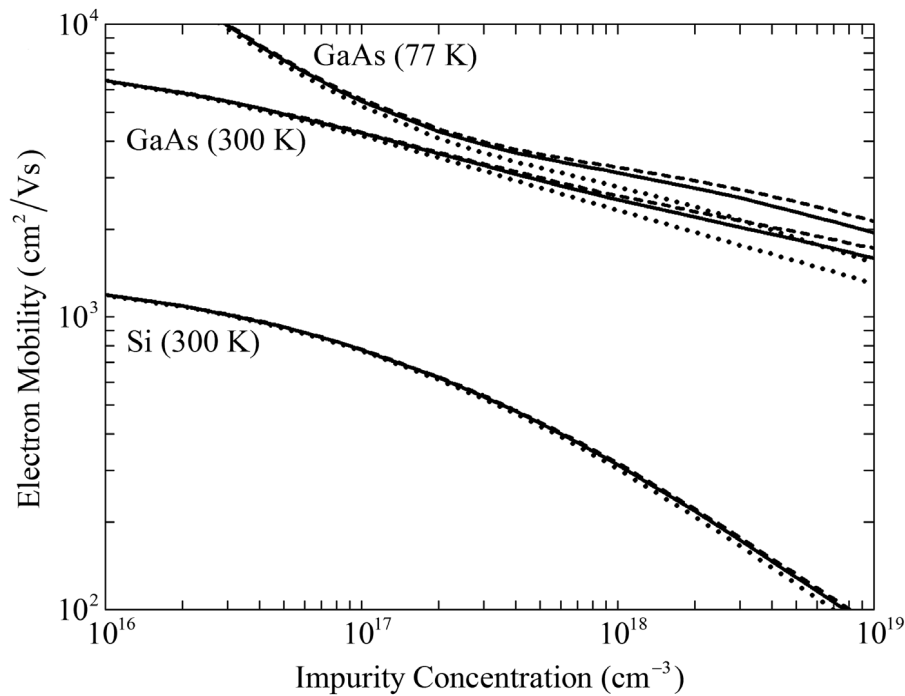

**Fig. 5** Low-field mobility of electrons in GaAs and Si calculated according to PM for different values of $\gamma$ (the rest calculation conditions are the same as in Figs. 2–4): $\gamma = 1$ (dotted curves), $\gamma = \gamma_0$ (solid curves) and $\gamma = 2^{1/2}$ (dashed curves).

One can conclude from the analysis of the obtained results represented in Figs. 2–4 that among the compared models (CWM, BHM and PM) describing the scattering of electrons by impurity ions PM is the best one since the calculated dependencies of the electron mobility on the donor impurity concentration in the semiconductors are in excellent agreement with the experimental data in all three figures.

# 4 Conclusion

Thus in the present work the new effective applied model describing the scattering of electrons by ionized impurity atoms in semiconductors is proposed. In comparison with such most frequently used applied models as the Conwel–Weisskopf model, the Brooks–Herring model and the Ridley one the application of the proposed model allows the best agreement between the theoretical calculations and experimental measurements of the electron mobility in semiconductors to be achieved.

# Acknowledgment

The author is very grateful to Springer for publication of the manuscript in "Journal of Computational Electronics". The final publication is available at http://link.springer.com/.
DOI 10.1007/s10825-013-0538-8